%% file: prl_D0toKpi9.tex
\documentclass[twocolumn,showpacs,aps,prl,superscriptaddress,preprintnumbers,amsmath,amssymb,floatfix]{revtex4}

\usepackage{graphicx}    
\usepackage{dcolumn}     
\usepackage{bm}          

\newcommand{\BABARPubYear}    {07}
\newcommand{\BABARPubNumber}  {011}

\newcommand{\SLACPubNumber}   {12463}

\input babarsym

\long\def\inst#1{\par\nobreak\kern 4pt\nobreak
    {\it #1}\par\vskip 10pt plus 3pt minus 3pt}

\RequirePackage{xspace}

\usepackage{relsize}
\def\babar{\mbox{\slshape B\kern-0.1em{\smaller A}\kern-0.1em
    B\kern-0.1em{\smaller A\kern-0.2em R}}}

\def\en         {\ensuremath{e^-}\xspace}   
\def\ep         {\ensuremath{e^+}\xspace}
 
\def\epem       {\ensuremath{e^+e^-}\xspace}

\def\mup        {\ensuremath{\mu^+}\xspace}
\def\mun        {\ensuremath{\mu^-}\xspace} 
\def\mumu       {\ensuremath{\mu^+\mu^-}\xspace}

\def\taup       {\ensuremath{\tau^+}\xspace}
\def\taum       {\ensuremath{\tau^-}\xspace}

\def\ellm       {\ensuremath{\ell^-}\xspace}
\def\ellp       {\ensuremath{\ell^+}\xspace}
\def\ellpm      {\ensuremath{\ell^{\pm}}\xspace}  
\def\ellmp      {\ensuremath{\ell^{\mp}}\xspace}  

\def\ellell     {\ensuremath{\ell^+ \ell^-}\xspace}
\def\nub        {\ensuremath{\overline{\nu}}\xspace}
\def\nunub      {\ensuremath{\nu{\overline{\nu}}}\xspace}
\def\nub        {\ensuremath{\overline{\nu}}\xspace}
\def\nunub      {\ensuremath{\nu{\overline{\nu}}}\xspace}
\def\nue        {\ensuremath{\nu_e}\xspace}
\def\nueb       {\ensuremath{\nub_e}\xspace}

\def\num        {\ensuremath{\nu_\mu}\xspace}
\def\numb       {\ensuremath{\nub_\mu}\xspace}

\def\nut        {\ensuremath{\nu_\tau}\xspace}
\def\nutb       {\ensuremath{\nub_\tau}\xspace}

\def\nul        {\ensuremath{\nu_\ell}\xspace}
\def\nulb       {\ensuremath{\nub_\ell}\xspace}

\def\g     {\ensuremath{\gamma}\xspace}
\def\gaga  {\ensuremath{\gamma\gamma}\xspace}  
\def\ggstar{\ensuremath{\gamma\gamma^*}\xspace}

\def\piz   {\ensuremath{\pi^0}\xspace}

\def\ppz   {\ensuremath{\pi^0\pi^0}\xspace}
\def\pip   {\ensuremath{\pi^+}\xspace}
\def\pim   {\ensuremath{\pi^-}\xspace}
\def\pipi  {\ensuremath{\pi^+\pi^-}\xspace}

\def\pimp  {\ensuremath{\pi^\mp}\xspace}
\def\kaon  {\ensuremath{K}\xspace}
\def\Kbar  {\kern 0.2em\overline{\kern -0.2em K}{}\xspace}

\def\Kz    {\ensuremath{K^0}\xspace}
\def\Kzb   {\ensuremath{\Kbar^0}\xspace}
\def\KzKzb {\ensuremath{\Kz \kern -0.16em \Kzb}\xspace}
\def\Kp    {\ensuremath{K^+}\xspace}
\def\Km    {\ensuremath{K^-}\xspace}
\def\Kpm   {\ensuremath{K^\pm}\xspace}

\def\KpKm  {\ensuremath{\Kp \kern -0.16em \Km}\xspace}
\def\KS    {\ensuremath{K^0_{\scriptscriptstyle S}}\xspace} 
\def\KL    {\ensuremath{K^0_{\scriptscriptstyle L}}\xspace} 
\def\Kstarz  {\ensuremath{K^{*0}}\xspace}

\def\Kstar   {\ensuremath{K^*}\xspace}

\def\Kstarp  {\ensuremath{K^{*+}}\xspace}

\def\Dbar    {\kern 0.2em\overline{\kern -0.2em D}{}\xspace}
\def\Db      {\ensuremath{\Dbar}\xspace}
\def\Dz      {\ensuremath{D^0}\xspace}
\def\Dzb     {\ensuremath{\Dbar^0}\xspace}
\def\DzDzb   {\ensuremath{\Dz {\kern -0.16em \Dzb}}\xspace}
\def\Dp      {\ensuremath{D^+}\xspace}
\def\Dm      {\ensuremath{D^-}\xspace}

\def\DpDm    {\ensuremath{\Dp {\kern -0.16em \Dm}}\xspace}
\def\Dstar   {\ensuremath{D^*}\xspace}

\def\Dstarz  {\ensuremath{D^{*0}}\xspace}

\def\Dstarp  {\ensuremath{D^{*+}}\xspace}
\def\Dstarm  {\ensuremath{D^{*-}}\xspace}

\def\B       {\ensuremath{B}\xspace}
\def\Bbar    {\kern 0.18em\overline{\kern -0.18em B}{}\xspace}
\def\Bb      {\ensuremath{\Bbar}\xspace}
\def\BB      {\ensuremath{\B {\kern -0.16em \Bb}}\xspace}
\def\Bz      {\ensuremath{B^0}\xspace}
\def\Bzb     {\ensuremath{\Bbar^0}\xspace}
\def\BzBzb   {\ensuremath{\Bz {\kern -0.16em \Bzb}}\xspace}

\def\Bu      {\ensuremath{B^+}\xspace}
\def\Bub     {\ensuremath{B^-}\xspace}

\def\Bpm     {\ensuremath{B^\pm}\xspace}

\def\BpBm    {\ensuremath{\Bu {\kern -0.16em \Bub}}\xspace}

\def\BorBbar    {\kern 0.18em\optbar{\kern -0.18em B}{}\xspace}
\def\DorDbar    {\kern 0.18em\optbar{\kern -0.18em D}{}\xspace}
\def\KorKbar    {\kern 0.18em\optbar{\kern -0.18em K}{}\xspace}

\def\jpsi     {\ensuremath{{J\mskip -3mu/\mskip -2mu\psi\mskip 2mu}}\xspace}

\mathchardef\Upsilon="7107
\def\Y#1S{\ensuremath{\Upsilon{(#1S)}}\xspace}

\def\FourS {\Y4S}

\mathchardef\Deltares="7101
\mathchardef\Xi="7104
\mathchardef\Lambda="7103
\mathchardef\Sigma="7106
\mathchardef\Omega="710A

\def\Deltabar{\kern 0.25em\overline{\kern -0.25em \Deltares}{}\xspace}
\def\Lbar{\kern 0.2em\overline{\kern -0.2em\Lambda\kern 0.05em}\kern-0.05em{}\xspace}
\def\Sigbar{\kern 0.2em\overline{\kern -0.2em \Sigma}{}\xspace}
\def\Xibar{\kern 0.2em\overline{\kern -0.2em \Xi}{}\xspace}
\def\Obar{\kern 0.2em\overline{\kern -0.2em \Omega}{}\xspace}
\def\Nbar{\kern 0.2em\overline{\kern -0.2em N}{}\xspace}
\def\Xb{\kern 0.2em\overline{\kern -0.2em X}{}\xspace}
\def\X {\ensuremath{X}\xspace}

\def\bpsikl     {\ensuremath{\Bz \to \jpsi \KL}\xspace}

\def\Bzbtox     {\ensuremath{\Bzb \to \X}\xspace}

\def\invfb   {\ensuremath{\mbox{\,fb}^{-1}}\xspace}

\def\mus  {\ensuremath{\rm \,\mus}\xspace}

\def\mus        {\ensuremath{\,\mu{\rm s}}\xspace}    

\def\degrees{\ensuremath{^{\circ}}\xspace}

\def\psoft {\ensuremath{\pi^+_s}\xspace}    
\def\psoftpm {\ensuremath{\pi^{\pm}_s}\xspace}  

\def\deltam {\ensuremath{\Delta M}\xspace}

\def\to                 {\ensuremath{\rightarrow}\xspace}
\def\pep2{PEP-II}
\def\BF{$B$ Factory}

\def\gsim{{~\raise.15em\hbox{$>$}\kern-.85em
          \lower.35em\hbox{$\sim$}~}\xspace}
\def\lsim{{~\raise.15em\hbox{$<$}\kern-.85em
          \lower.35em\hbox{$\sim$}~}\xspace}

\def\epstag  {\ensuremath{\varepsilon_{\rm tag}}\xspace}

\def\jetset74   {\mbox{\tt Jetset \hspace{-0.5em}7.\hspace{-0.2em}4}\xspace}

\usepackage{relsize}

\def\Mnu{\ensuremath{{\cal M}_\nu^2}}

\begin{document}

\begin{flushleft}
{\babar-PUB-\BABARPubYear/\BABARPubNumber \\
SLAC-PUB-\SLACPubNumber}     \\[1mm]
\end{flushleft}

\title{Measurement of the Absolute Branching Fraction of
{\boldmath{$\Dz \rightarrow \Km \pip$}}}


\input{authors_final.tex}

\date{\today} 

\begin{abstract}
We measure the absolute branching fraction for $\Dz \rightarrow \Km \pip$ 
using partial reconstruction of $\Bzb \rightarrow D^{*+} X \ell^{-}
\bar{\nu}_{\ell}$ decays, in which
only the charged lepton and the pion from the decay $D^{*+} \rightarrow D^{0} \pi^{+}$ are used.
Based on a data sample of 230 million \BB pairs collected at the $\Upsilon(4S)$
resonance with the \babar\ detector at the PEP-II asymmetric-energy $B$
Factory at SLAC, we obtain  ${\cal B}(\Dz \rightarrow \Km \pip) = 
(4.007 \pm 0.037 \pm 0.072)\%$, where the first uncertainty is statistical and the second 
is systematic. 
\end{abstract}


\maketitle

The decay $\Dz \rightarrow \Km \pip$~\cite{conjugate} is a reference mode for 
the measurements of the branching fractions of the $D^0$ to any other final state.
A precise measurement of the value of ${\cal B}(\Dz \rightarrow \Km \pip)$ 
improves our knowledge of most of the decays of the $B$ mesons, and of fundamental 
parameters of the Standard Model. For instance, the largest systematic uncertainty on 
the branching ratio ${\cal B}(B^0 \rightarrow \dsm \ellp \nu_\ell)$, and the  
experimental uncertainty on the determination of the Cabibbo-Kobayashi-Maskawa matrix 
element V$_{cb}$ from that semileptonic decay are induced by the uncertainty
on  ${\cal B}(\Dz \rightarrow \Km \pip)$.

CLEO-c~\cite{cleoc05} has recently published the most precise result on this branching fraction, 
which is widely used~\cite{pdg06}. 
We present here a more precise measurement based on a different technique.
We identify $\Dz \rightarrow \Km \pip$ decays in a sample of $\Dz$ mesons 
from $\dsp \rightarrow \Dz \pip$ decays 
and obtained with partial reconstruction of $\Bzb \rightarrow D^{*+} (X) \ell^{-} \bar{\nu}_{\ell}$.

The data sample used in this analysis consists of an integrated luminosity of 210\invfb, 
corresponding to 230 million \BB pairs, collected at the $\Y4S$ resonance (on-resonance) 
and 22\invfb collected 40\mev below the resonance (off-resonance) by the \babar\ detector. 
The off-resonance events are used to subtract the non-\BB\ (continuum) background.
A simulated sample of $\BB$ events with integrated luminosity equivalent to approximately 
five times the size of the data sample is used for efficiency computation and background studies.

A detailed description of the \babar\ detector is provided
elsewhere~\cite{babar_nim}. High-momentum particles are reconstructed by matching 
hits in the silicon vertex tracker (SVT) with track elements in the drift chamber (DCH). 
Lower momentum tracks, which do not leave signals on many wires in the DCH due 
to the bending induced by the 1.5 T solenoid field, are reconstructed solely in the SVT.
Charged hadron identification is performed by combining the  measurements of 
the energy deposition in the SVT and in the DCH with the information from a 
Cherenkov detector (DIRC).
Electrons are identified by the ratio of
the energy deposited in the calorimeter (EMC)  to the track momentum, the transverse
profile of the shower, the energy loss in the DCH, and 
the Cherenkov angle in the DIRC.
Muons are identified in the instrumented flux return (IFR), composed
of resistive plate chambers and layers of iron.

We preselect a sample of hadronic events with at least four charged tracks.
To reduce continuum background, we require that the ratio of the 2$^{nd}$ to the
0$^{th}$ order Fox-Wolfram~\cite{wolfram} variables be less than 0.6.
We then select a sample of partially reconstructed $B$ mesons in the channel 
$\Bzb \rightarrow D^{*+} (X) \ell^{-} \bar{\nu}_{\ell}$, 
by retaining events containing a charged lepton ($\ell = e,\,\mu$) and a low momentum 
pion (soft pion, $\pi^+_{s}$) which may arise from the decay $D^{*+}\to \Dz \pi^+_{s}$.
This sample of events is referred to as the ``inclusive sample''.
The lepton momentum \cite{frame}  must be in the range $1.4 < p_{\ellm} < 2.3 \gevc$ and 
the soft pion candidate must satisfy $60 < p_{\pi^{+}_{s}} < 190 \mevc$. 
The lepton and soft pion minimum momenta are optimized to minimize
uncertainties due to charm production in $B$ decays and tracking errors, 
respectively. Maximum momentum selections are determined by the available
phase space. The two tracks must be consistent with originating from a common vertex, 
constrained to the beam-spot in the plane transverse to the beam axis. Then we
combine $p_{\ellm}$, $p_{\pi^{+}_{s}}$ 
and the probability from the vertex fit into a likelihood ratio variable, 
optimized to reject \BB\ background. 
Using conservation of momentum and energy, 
the invariant mass squared of the undetected neutrino is calculated as
\begin{eqnarray}
\Mnu \equiv (E_{\mbox{\rm \small beam}}-E_{{D^*}} - 
E_{\ell})^2-({\vec{p}}_{{D^*}} + {\vec{p}}_{\ell})^2 ,
\label{eqn:mms}
\end{eqnarray}
where $E_{\mbox{\rm \small beam}}$ is half the total center-of-mass energy and $E_{\ell}~(E_{{D^*}})$ 
and ${\vec{p}}_{\ell}~({\vec{p}}_{{D^*}})$ are the energy and momentum 
of the lepton (the $D^*$ meson).  Since the magnitude of the $B$ meson
momentum, $p_{B}$, is sufficiently 
small compared to $p_{\ell}$ and $p_{D^*}$, we set
$p_{B}$ = 0
 in obtaining Eq.~\ref{eqn:mms}.
As a consequence of the limited phase space available in the $D^{*+}$
decay, the soft pion is emitted nearly at rest in the $D^{*+}$ rest frame.
The $D^{*+}$ four-momentum can therefore be computed by approximating 
its direction as that of the soft pion, and parameterizing its momentum as 
a linear function of the soft-pion momentum.
We select pairs of tracks with opposite electric charge for our signal (\ellmp \psoftpm)
and same-charge pairs (\ellpm \psoftpm) for background studies.  

All events where $\dsp$ and $\ell^-$ originate from the same $B$-meson, producing 
a peak near zero in the \Mnu\ distribution, are considered as signal candidates.
Several processes contribute:
(a) $\Bzb \rightarrow D^{*+} \ell^{-} \bar\nu_{\ell}$ decays (primary);
(b) $\Bb \rightarrow D^{*+} (\mathrm{n}\pi) \ell^- \bar{\nu}_{\ell}$ where the $D^{*+} (\mathrm{n}\pi)$ may 
or may not originate from an excited charm state (\dstrstr) and n$ \ge 1$;
(c) $\Bzb \rightarrow D^{*+}\Db $, $\Db \rightarrow \ell^{-}X$ and 
$\Bzb \rightarrow D^{*+}\tau^- \bar{\nu}_{\tau} $, $\tau^- \rightarrow
\ell^{-}\bar{\nu}_{\ell}\nu_{\tau} $
(cascade); 
(d) $\Bzb \rightarrow D^{*+} h^-$ (fake-lepton), where the hadron ($h = \pi,K)$ is erroneously 
identified as a lepton (in most of the cases, a muon). We also include radiative events, where photons 
with energy above 1 MeV are emitted by any charged particle using PHOTOS v2.03~\cite{photos}. 
The signal region is $\Mnu > -2~$GeV$^{2}/c^4$ and the sideband is
$-10 < \Mnu < -4~$GeV$^{2}/c^4$.

\begin{figure}[!htb]
\begin{center}
\includegraphics[width=9cm]{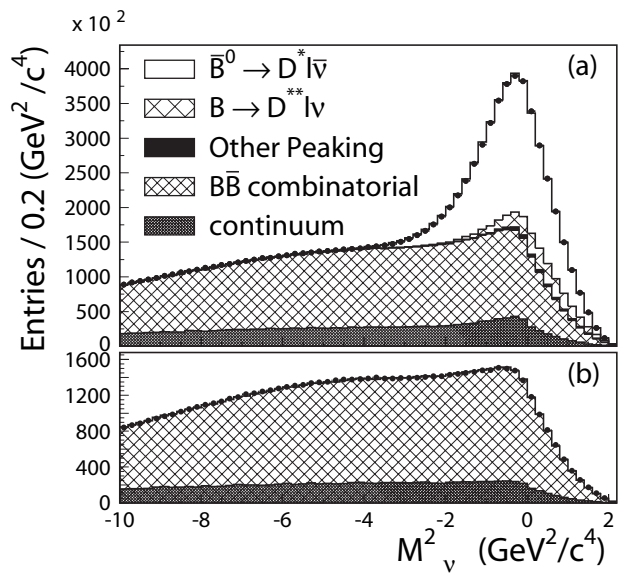}
\caption{The ${\cal M}_\nu^{\,2}$ distribution of the inclusive
  sample, for right-charge (a) and wrong-charge (b) samples.
The data are represented by solid points with uncertainty.
The MC fit results are overlaid to the data, as explained in the figure.}
\label{fig:incl_yield}
\end{center}
\end{figure}

The background in the inclusive sample consists of continuum and
combinatorial \BB\ events, which also 
include events where true \dsp\ and  $\ell^-$ from the two different $B$ mesons are combined.
We determine the number of signal events in our sample with a minimum
$\chi^2$ fit to the \Mnu\ distribution in the 
interval $-10 <\Mnu< 2.5~$GeV$^2$/c$^4$. 
We perform the fit in ten bins of the lepton momentum
in order to reduce the sensitivity of the result to the details of the simulation.
In each bin we fix the continuum contribution to the off-resonance events,
rescaled to account for the luminosity ratio between the on- and the
off-resonance samples, while we vary independently the number of 
signal events from primary, from \dstrstr, and from combinatorial \BB,
assuming the shapes predicted by the simulation.
We fix the contributions from cascade and fake-lepton decays, which account for about 3\% of the signal sample, 
to the Monte Carlo (MC) prediction. 
We fit eight different sets, divided by lepton kind and run
condition. The reduced $\chi^2$s range between 1.1 and 1.4.
Figure~\ref{fig:incl_yield}(a) shows the result of the fit in the \Mnu\ projection. 
The number of signal events with $\Mnu > -2$ GeV$^2$/c$^4$ is 
$N^{\rm incl} = (2170.64 \pm 3.04 (stat) \pm 18.1 (syst)) \times 10^3$. The statistical
uncertainty includes the statistical uncertainties of the off-resonance and of the simulated events.

We then reconstruct for $\Dz \to K^- \pip$ decays in the inclusive sample.
We consider all tracks in the event, aside from the \ellm
and \psoft, with momenta in the direction transverse to
the beam axis exceeding 0.2 \gevc.
We combine pairs of tracks with opposite charge, and compute the
invariant mass $M_{K\pi}$ assigning the kaon mass to the track with
charge opposite the $\pi_s$ charge.
The kaon candidate must satisfy a loose kaon identification criterion
that 
retains more than 80\% of true kaons, while rejecting more than 95\% of pions.
We select events in the mass range $1.82 < M_{K\pi} < 1.91$ \gevcc. 
We combine each \Dz\ candidate with the \psoft and compute the mass
difference $\Delta M = M(\Km \pip \psoft) - M(\Km \pip)$. We look for
signal in the range of $142.4 < \Delta M < 149.9$ \mevcc. 

This exclusive sample consists of signal events and of the following background sources:
continuum, combinatorial \BB, uncorrelated peaking \dsp\ and Cabibbo-suppressed decays.
We subtract the continuum background using rescaled off-resonance events selected with 
the same criteria as the on-resonance data. Combinatorial events are due to any combination 
of three tracks, in which at least one does not come from the \dsp. We determine their number 
from simulated \BB events. We normalize the simulated events to the data in the \deltam\ sideband, 
$153.5 < \Delta M < 162.5$ \mevcc, properly accounting for the
small fraction of signal events (less than 1\%) contained in the sideband.
We verify that the background shape is properly described in the
simulation using a sample of \dsp -depleted
events, obtained as follows. 
We use wrong-charge events where the kaon has the same charge as the
$\pi_s$, selected in the \Mnu\ sideband.
More than 95\% of the events so selected
in the $\Delta M$ signal region are combinatorial background, with a residual peaking 
component from Cabibbo suppressed decays 
($\Kp\Km$ and \pip \pim, see below). After normalizing the level of
the simulated events in the sideband, 
the number of events in the signal region is consistent with the data within 
the statistical precision of $\pm 1.3\%$.

The background from uncorrelated peaking \dsp\ decays occurs when the \dsp\ and the \ellm\ 
originate from the two different $B$ mesons. These events exhibit a peak in \deltam\ but behave 
as combinatorial background in \Mnu . We compute their number in the \Mnu\ sideband data and 
rescale it to the \Mnu\ signal region using the \Mnu\ distribution of
the combinatorial simulated events. 

\begin{figure}[!htb]   
\begin{center}
\vspace*{-2.2cm}
\includegraphics[width=9cm]{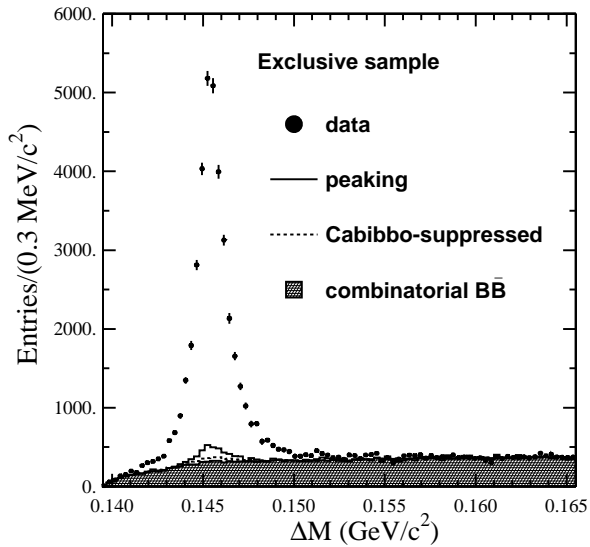}
\vspace*{-2cm}
\caption{Continuum subtracted $\Delta {M}$ distribution for data (points with error bars) and 
backgrounds overlaid as explained in the figure.}
\label{fig:excl_yield}
\end{center}
\end{figure}

Cabibbo-suppressed decays  $\Dz \to \Km\Kp$ ($\pim \pip$) contribute to the peaking background, 
where one of the kaons (pions) is wrongly identified as a pion (kaon). Simulation shows that 
these events peak in \deltam, while they exhibit a broad $M_{K\pi}$ distribution. 
We subtract this background source using the simulation prediction.
It should be noted that the contribution from doubly Cabibbo suppressed decays is negligible.
Figure~\ref{fig:excl_yield} shows the continuum-subtracted distribution for the data with 
the simulated \BB\ backgrounds overlaid. 

The exclusive selection yields $N^{\rm excl} = (3.381 \pm 0.029) \times 10^4$ signal events, where 
the uncertainty is statistical only. The detailed composition of the inclusive and exclusive data sets is 
listed in Table~\ref{tab:excl_results}.
\begin{table}[!htb]
\caption {The inclusive and exclusive samples.}
\begin{center}
\begin{tabular}{lrr} \hline \hline
Source              & Inclusive ($\times 10^6$)  &  Exclusive ($\times 10^4$)
\\ \hline 
Data                & $4.4124\pm 0.0021$  & $4.727 \pm 0.022$   \\ 
Continuum           & $0.46  \pm 0.0021$  & $0.309 \pm 0.017$   \\
Combinatorial \BB\  & $1.7817\pm 0.0007$  & $0.819 \pm 0.005$   \\       
Peaking             & --                  & $0.163 \pm 0.008$   \\
Cabibbo-suppressed  & --                  & $0.055 \pm 0.001$   \\ \hline 
Signal              & $2.1706 \pm 0.0030$ & $3.381 \pm 0.029$   \\ \hline  \hline 
\end{tabular}
\end{center}
\label{tab:excl_results}
\end{table}

We compute the branching fraction as
\begin{eqnarray}
{\cal B}(\Dz \rightarrow \Km \pip) & = &  
             {N^{\rm excl}/(N^{\rm incl}}\zeta \varepsilon_{(\Km \pip)}),
\label{eq:branching_ratio}
\end{eqnarray}
where $\varepsilon_{(\Km \pip)} = (36.96 \pm 0.09)\%$ is the $\Dz$
reconstruction efficiency from simulation, and 
$\zeta = 1.033 \pm 0.002$ is the selection bias introduced by the partial reconstruction.
Only the statistical uncertainties are reported here.  
The bias factor $\zeta$ accounts for the larger efficiency of the inclusive 
event reconstruction for final states with two or fewer tracks from
\Dz\ decays due to the smaller density of hits 
near the \psoft\ track. We study these effects by comparing data and simulated
distributions of the number of charged tracks in each event
($n_{\rm trk}$) and of other quantities sensitive to the soft
pion isolation (angle to nearest track and track density within $10^{\circ}$
cone around the \psoft\ direction). We weight simulated events to reproduce 
the data and recompute the bias. We observe an efficiency
variation of $0.33\%$ due to $n_{\rm trk}$ and $0.08\%$  due to the other
variables. The bias does not depend on some other variables
($p_{\psoft}$, number of \psoft\ hits in the SVT).
The systematic uncertainty due to this selection is $\pm 0.35\%$.

The main systematic uncertainty on $N^{\rm incl}$ is due to 
the non-peaking combinatorial \BB\ background. 
We perform the same fit to the $\ellpm \psoftpm$ background control sample 
and the signal-dominated sample. We take the systematic uncertainty in the combinatorial 
background to be the RMS scatter in the ratio, calculated for each \Mnu\ bin 
as shown in Figure~\ref{fig:incl_yield}(b), of continuum-subtracted data to 
the value of the combinatorial background determined from the fit, resulting 
in an uncertainty of 0.89\%.
As first noticed in~\cite{cleo98}, the decays $\Bzb \to \ell^- \bar{\nu_\ell} D^+$, with 
$D^+ \to K^* \rho (\omega) \pi^+$, constitute a right-charge peaking background, because the charged pion 
is produced almost at rest in the \Dp\ rest frame.
In order to estimate the systematic uncertainty due to this peaking combinatorial background, we vary its total 
fraction by $\pm 100\%$ in the \BB\ events in the MC. The corresponding systematic 
uncertainty is $\pm 0.34\%$.

We consider systematic uncertainties affecting the signal
\Mnu\ distribution. Final state photon radiation in \Dz\ decays alters the distribution of
$M_{\Km \pip}$ and thus affects the efficiency computation. We estimate a
systematic uncertainty of $\pm 0.50\%$ due to the final state photon radiation 
by varying by $\pm 30\%$ the fraction of reconstructed events in the simulation 
where at least one photon above 1 MeV is emitted in the $\Dz \rightarrow \Km \pip$ decay. 

We also vary, by $\pm 30\%$, the fractions of cascade and fake-lepton decays, 
which are not determined by the fit. Finally, we vary in turn by $\pm 100\%$ the number of events 
from each of the five sources constituting the \dstrstr\ samples (two narrow and two broad resonant 
states, and non-resonant $D^{*+}\pi$ combinations; these last are
described using the model of ref. \cite{Goy}). We repeat 
the measurement and take the variation as the systematic uncertainty.

The dominant contribution to the systematic uncertainty on $N^{\rm excl}$
is due to the charged-track reconstruction efficiency. 
The single charged-track reconstruction efficiency is determined with
$0.50\%$ precision, which corresponds to $\pm 1.00\%$ overall uncertainty. 
The efficiency for $K^-$ identification is measured with 
$\pm 0.70\%$ systematic uncertainty from a large sample of $\dsp \to
\Dz\pi_s^+, \Dz \to K^- \pi^+$ decays,
 produced in $\epem \rightarrow c \bar{c}$ events.
To estimate the systematic uncertainty due to the combinatoric
background subtraction on $N^{\rm excl}$, 
we first vary the number of events from combinatorial background below the signal peak by $\pm
1.3\%$, corresponding to the statistical uncertainty obtained from the
control sample described above. This translates in $\pm 0.3\%$ systematic uncertainty on the result. 
We vary the number of signal events contained in the sideband by $\pm 30\%$ for background
normalization. This induces a systematic uncertainty of $\pm 0.16
\%$. We vary the fraction of events from Cabibbo suppressed decays by $\pm 10\%$.  
The systematic uncertainty due to the uncorrelated peaking (from data) is negligible.

When comparing the simulated $M_{K\pi}$ distribution to the data in a
high purity signal sample
(obtained by asking, in addition to the other cuts, that the hard pion
fails K, p and $\ell$ identification criteria and 
$0.1435< \Delta M< 0.1475 \gevcc$) we observe a slight discrepancy, causing $\pm 0.56\%$ systematic
uncertainty in the reconstruction efficiency. 
We compute the total relative systematic uncertainty of $\pm 1.80\%$ from the quadratic sum of all 
uncertainties described above and listed in Table~\ref{tab:sys_errors}. 
\begin{table}[!htb]
\caption{The systematic uncertainties of ${\cal B}(\Dz \rightarrow \Km \pip)$.}
\begin{center}
\begin{tabular}{llr} \hline \hline
                       & Source                                   & $\delta({\cal B})/{\cal B}$ (\%)  \\ \hline
                       & Selection bias                           & $\pm 0.35$        \\  \hline
$N^{\rm incl}$         & Non-peaking combinatorial background     & $\pm 0.89$        \\
                       & Peaking combinatorial background         & $\pm 0.34$        \\
                       & Soft pion decays in flight               & $\pm 0.10$        \\ 
                       & Fake leptons                             & $\pm 0.08$        \\
                       & Cascade decays                           & $\pm 0.08$        \\
                       & Monte Carlo events shape                 & $\pm 0.08$        \\
                       & Continuum background                     & $\pm 0.05$        \\
                       & \dstrstr\ production                     & $\pm 0.02$        \\
                       & Photon radiation                         & $\pm 0.02$        \\  \hline
$N^{\rm excl}$         & Tracking efficiency                      & $\pm 1.00$        \\
                       & $K^-$ identification                     & $\pm 0.70$        \\  
                       & \Dz invariant mass                       & $\pm 0.56$        \\
                       & Photon radiation in \Dz\ decay           & $\pm 0.50$        \\ 
                       & Combinatorial background shape           & $\pm 0.30$        \\
                       & Combinatorial background normalization   & $\pm 0.16$        \\
                       & Soft pion decay                          & $\pm 0.12$        \\
                       & Cabibbo-suppressed decays                & $\pm 0.10$        \\  \hline 
Total                  &                                          & $\pm 1.80$        \\  \hline \hline   
\end{tabular}
\end{center}
\label{tab:sys_errors}
\end{table}   
We cross check our results using different definitions of the
$\Delta M$ and $M_{\Km\pip}$ signal regions and particle
identification. We split our data into different sub-samples,
depending on the run conditions.  
All the results are consistent.

In summary, we have measured the absolute branching fraction of $\Dz \rightarrow \Km \pip$ decay 
with partial reconstruction of $\Bzb \rightarrow D^{*+} (X) \ell^{-}
\bar\nu_{\ell}$, and obtain the result
\begin{eqnarray}
{\cal B}(\Dz \rightarrow \Km \pip) & = & (4.007 \pm 0.037 \pm 0.072)\% ,
\label{eq:final_result}
\end{eqnarray}
where the first uncertainty is statistical and the second uncertainty is
systematic. This result is comparable in precision with
the present world average, and it is consistent with it within two
standard deviations.

\begin{acknowledgments}
\input{acknow_PRL_final}

\end{acknowledgments} 


\end{document}

%% file: authors_final.tex
%
\author{B.~Aubert}
\author{M.~Bona}
\author{D.~Boutigny}
\author{Y.~Karyotakis}
\author{J.~P.~Lees}
\author{V.~Poireau}
\author{X.~Prudent}
\author{V.~Tisserand}
\author{A.~Zghiche}
\affiliation{Laboratoire de Physique des Particules, IN2P3/CNRS et Universit\'e de Savoie, F-74941 Annecy-Le-Vieux, France }
\author{J.~Garra~Tico}
\author{E.~Grauges}
\affiliation{Universitat de Barcelona, Facultat de Fisica, Departament ECM, E-08028 Barcelona, Spain }
% ...
\collaboration{The \babar\ Collaboration}
\noaffiliation

%% file: acknow_PRL_final.tex
We are grateful for the excellent luminosity and machine conditions
provided by our \pep2\ colleagues, 
and for the substantial dedicated effort from
the computing organizations that support \babar.
The collaborating institutions wish to thank 
SLAC for its support and kind hospitality. 
This work is supported by
DOE
and NSF (USA),
NSERC (Canada),
CEA and
CNRS-IN2P3
(France),
BMBF and DFG
(Germany),
INFN (Italy),
FOM (The Netherlands),
NFR (Norway),
MIST (Russia),
MEC (Spain), and
PPARC (United Kingdom). 
Individuals have received support from the
Marie Curie EIF (European Union) and
the A.~P.~Sloan Foundation.

%% file: prl_D0toKpi9.bbl
\begin{thebibliography}{99}
\bibitem{conjugate}
The inclusion of  charge-conjugate reactions is implied throughout this article.
\bibitem{frame}
Four-momenta are computed in the \FourS\ rest frame.
\end{thebibliography}
